# Dynamics of Solvated Electrons during Femtosecond Laser-Induced Plasma Generation in Water

Noritaka Sakakibara,[1,2,*] Tsuyohito Ito,[1,2] Kazuo Terashima,[1,2] Yukiya Hakuta,[2] and Eisuke Miura[2,*]

[1] Department of Advanced Materials Science, Graduate School of Frontier Sciences, The University of Tokyo, 5-1-5 Kashiwanoha, Kashiwa, Chiba 277-8561, Japan

[2] AIST-UTokyo Advanced Operando-Measurement Technology Open Innovation Laboratory (OPERANDO-OIL), National Institute of Advanced Industrial Science and Technology (AIST), 5-1-5 Kashiwanoha, Kashiwa, Chiba, 277-8589, Japan

**Keywords:** solvated electron, femtosecond laser, breakdown, plasma, pump-probe measurement

[*] E-mail: n.sakakibara@plasma.k.u-tokyo.ac.jp, e-miura@aist.go.jp

**Abstract:** We studied the dynamics of solvated electrons in the early stage of plasma generation in water induced with an intense femtosecond laser pulse. According to the decay kinetics of solvated electrons, fast recombination process of solvated electrons (geminate recombination) occurred with a more prolonged lifetime (500 ps to 1 ns) than that observed in previous pulse photolysis studies (10–100 ps). This unusually longer lifetime is attributed to additional production of solvated electrons due to abundant free electrons generated with the laser-induced plasma, implying significant influence of free electrons on the dynamics of solvated electrons.

# I. INTRODUCTION

Femtosecond laser-induced plasma produced in liquid and aqueous media is leading to applications such as nanomaterials design [1–3] and cellular nanosurgery [4–6]. By tightly focusing an intense femtosecond laser pulse with an energy of 1 mJ or more [6], a high-density plasma with $10^{18}$–$10^{21}$ cm$^{-3}$ electron density is induced by multiphoton ionization, tunneling ionization, and collisional ionization [8, 9]. The abundant free electrons are simultaneously solvated to form solvated electrons ($e_{aq}^-$). Solvated electron is one of the important species that highly influences the chemical reactions [10, 11] and physical processes [8] in aqueous media through charge transfer and charge transport, as vigorously investigated in pulse photolysis studies using a low energy (10–100 μJ) femtosecond laser pulse [11–17]. However, there is no experimental investigation of the dynamics of solvated electrons during plasma generation by an intense femtosecond laser pulse. The dynamics of solvated electrons might be influenced and drastically changed by the abundant free electrons with laser-induced plasma.

In this study, for a better understanding of the early stage of plasma generation in water by an intense femtosecond laser pulse, we observed the dynamics of solvated electrons by a pump-probe technique using femtosecond laser pulses. Importantly, we found that annihilation of solvated electrons by fast recombination process with counter ions and radicals (geminate recombination) occurred with a longer lifetime than that observed in previous pulse photolysis studies, possibly due to the additional production of solvated electrons by abundant free electrons generated with laser-induced plasma.

## II. EXPERIMENT

Figure 1 depicts a pump-probe measurement setup. Fifty femtosecond laser pulses with 800 nm center wavelength were delivered from a Ti:sapphire laser system using chirped pulse amplification. The laser system consisted of a mode-locked oscillator, pulse stretcher, regenerative amplifier, multi-pass amplifier, and pulse compressor, and it was operated at the repetition rate of 1 Hz. One pulse, which was used as a pump pulse, was focused with an aspherical quartz lens with a focal length $f = 12.5$ mm into the middle of a quartz cell filled with purified water (electrical conductivity, 0.07 μS/cm). This pump pulse induced the generation of plasma in water. The energy of the pump pulse was regulated at 5.2±0.2 mJ right before it entered the lens. The other pulse, which was used as a probe pulse, illuminated the plasma in a direction perpendicular to that of the pump pulse propagation. The energy of the probe pulse was less than 0.2 mJ. The wavelength of the probe beam, which is 800 nm in this study, is suitable for monitoring solvated electrons because it is close to the peak wavelength (700 nm) of the absorption spectrum of solvated electrons [18]. Time-resolved observation was performed by controlling the optical delay between the pump and probe pulses by changing the optical length of the probe pulse using a micrometer. Furthermore, the time resolution of our setup was confirmed to be better than 1 ps, which is enough to monitor picosecond dynamics of solvated electrons. To verify the observation of solvated electrons, experiments were also performed in 0.2 M KNO$_3$$aq$ (21 mS/cm, pH = 6.8) and 0.4 M KNO$_3$$aq$ (38 mS/cm, pH = 6.2) instead of purified water because NO$_3^-$ is a well-established anionic scavenger of solvated electrons at various conditions, such as photolysis [19], radiolysis [20–22], photoelectron detachment [23], and plasma irradiation [24].

Using the pump-probe system, we performed time-resolved shadowgraph imaging, absorption measurement, and laser interferometry. For visualizing the plasma, shadowgraph images were recorded with a CCD camera through a lens with a focal length $f = 150$ mm. A biprism (with top angle of 177°) was inserted behind the lens during interferometry [25] for estimating the density of free electrons.

Shadowgraph and interferometry images could be captured with a single laser shot. By dividing a probe beam into two with a half mirror before it entered the cell, the intensities $I_{sig}$ and $I_{ref}$ were recorded using biplanar phototubes for the same laser shot, where $I_{sig}$ and $I_{ref}$ refer to the intensity of the probe beam that was and was not passed through the plasma, respectively. Absorption measurement was conducted by monitoring the $I_{sig}/I_{ref}$ ratio to determine the kinetics of solvated electrons. Although the absorption could be basically measured with a single laser shot in our method, the optical density was estimated from the average of 15 laser shots. In front of the biplanar phototubes, interference filters with a bandwidth of 10 nm at 800 nm were inserted to eliminate the stray light from the plasma. Further, to perform spatially resolved absorption measurement, the beam diameter of probe pulses was decreased to ~1.5 mm with an iris.

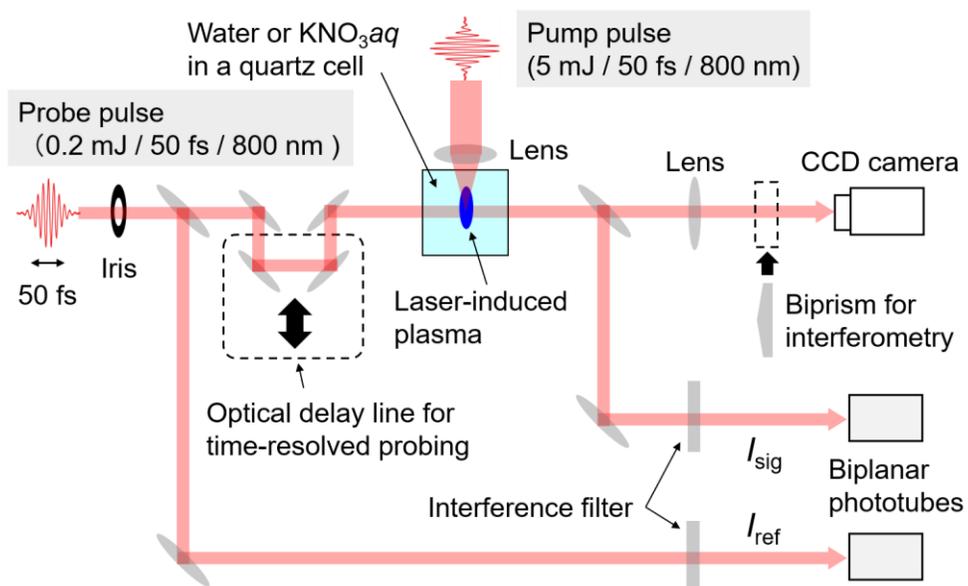

**FIG. 1.** Schematic of the pump-probe system for spatial and temporal measurement of the dynamics of solvated electrons. The biprism was installed on the laser path only when interferometry measurement was performed.

## III. RESULTS AND DISCUSSION

### A. Visualization of distribution of solvated electrons and plasma plume initiation

Rows (a), (b), and (c) in Fig. 2 show shadowgraph images in pure water, and 0.2 M and 0.4 M KNO$_3$ $aq$ for different probe delays, respectively. In Fig. 2, a pump pulse propagated from the right side of the image. The time at which a pump pulse reaches the focal point is defined as the start time of the probe delay $t = 0$ ps. The shadowgraph images in Fig. 2 reveal two types of shadows. One is a dilute shadow along the laser path that appears simultaneously with the formation of a focusing cone of the pump pulse at $t = 0$ ps. The other is a dense shadow in the vicinity of the focal point that grows over the delay time. Even with a delay of more than 400 ps, the shadows did not extend beyond the focal point because almost all the energy of the pump pulse might have been consumed in the generation of plasma plume and solvated electrons or scattered by the plasma itself. The dilute shadow faded in a few nanoseconds in pure water (Fig. 2(a)), whereas it faded rapidly within 400 ps and completely disappeared at 2.7 ns when 0.2 M or 0.4 M KNO$_3$ $aq$ were used as the media (Fig. 2(b) and (c)). Furthermore, the dilute shadow became thinner at higher KNO$_3$ concentration. These results indicate that the dilute shadow can be attributed to solvated electrons because NO$_3^-$ is a strong scavenger of solvated electrons [19, 23]. On the other hand, the dense shadow is considered to represent plasma plume initiation accompanied by abundant free electrons [26]. Therefore, shadowgraph imaging reveals the behavior of the laser-induced plasma; solvated electrons are widely distributed in the focusing cone of the pump laser pulse, while a plasma plume with a high electron density is initiated only in the vicinity of the focal point. The free and solvated electrons are initially produced by multiphoton absorption ionization process [8], and avalanche ionization caused by the collision of free electrons with surrounding water molecules induces further ionization and plasma plume initiation [8], when the laser intensity exceeds the breakdown threshold [7, 27, 28]. The laser intensity calculated from the pump pulse energy and the beam diameter at the edge of the plasma plume is $4 \times 10^{13}$ W/cm$^2$, which is similar to the breakdown threshold value reported previously for the femtosecond laser-induced breakdown in water [7]. All the measurements in this study were considered not to be disturbed by the temperature elevation of water and the expansion of cavitation bubble because the temperature elevation and the bubble expansion appear on slower timescale (10 ns or later) [5, 26, 27]. In other words, this study observed the dynamics of laser-induced plasma and solvated electrons before the expansion of plasma.

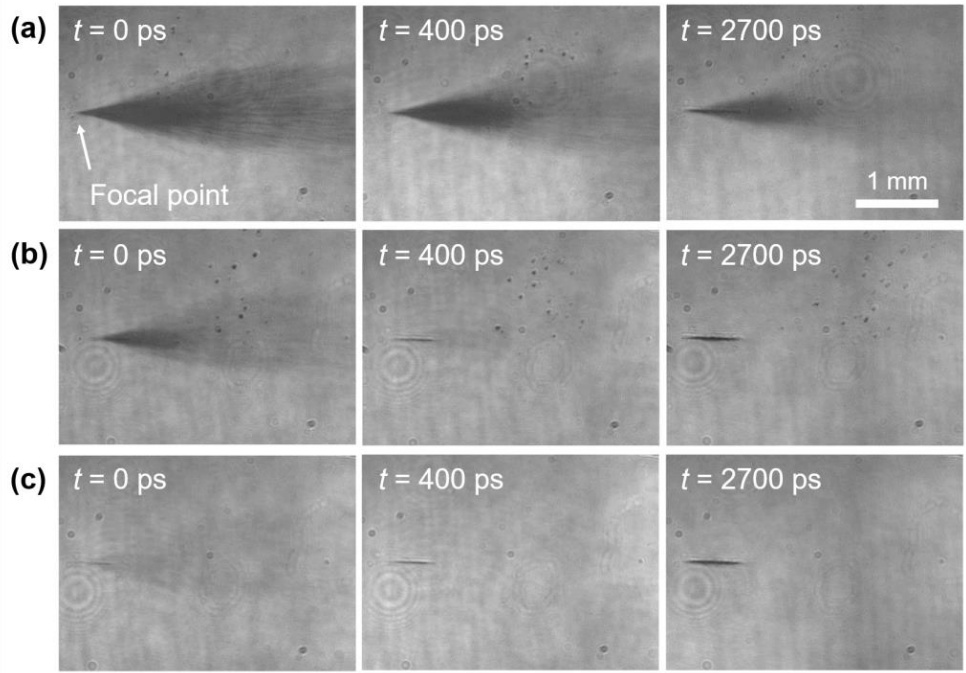

**FIG. 2.** Shadowgraph images of the laser-induced plasma in (**a**) pure water, (**b**) 0.2 M $KNO_3 aq$, and (**c**) 0.4 M $KNO_3 aq$ at different probe delays. The small dotted shadows are microbubbles generated by the laser-induced breakdown.

## B. Decay kinetics of solvated electrons

To investigate the kinetics of solvated electrons, laser absorptions in the vicinity of the focal point (probe area A in Fig. 3) and in the middle of the focusing cone (probe area B in Fig. 3) were measured. As shown in Fig. 3, temporal change in the optical density was obtained at each probe area. In this study, the optical density was calculated using Equation (1), from which the density of solvated electrons was estimated using the Lambert-Beer's law [29],

$$Optical\ density = -log\left(\frac{I_{sig}-(1-\alpha)I_{ref}}{\alpha I_{ref}}\right) = \varepsilon[e_{aq}^-]L \qquad (1)$$

where $\alpha$ is ratio of the shadow area within the probe beam at $t = 0$ ps to the whole probe beam area ($\alpha = 0.25$ at probe area A and $\alpha = 1$ at probe area B), $\varepsilon$ is the extinction coefficient of solvated electrons (19000 $M^{-1}cm^{-1}$) at the wavelength of 800 nm [18], $[e_{aq}^-]$ is the spatially averaged density of solvated electrons in the shadow within the probe beam, and $L$ is the path length of the probe pulse in the plasma. $L$ was determined from the transverse size of the shadowgraph images, assuming axial symmetry. For example, the density of solvated electrons at $t = 0$ ps in pure water was estimated to be $8.8\times10^{-4}$ M ($5.3\times10^{17}$ $cm^{-3}$) at probe area A with $L = 0.41$ mm, and $6.3\times10^{-5}$ M ($3.8\times10^{16}$ $cm^{-3}$) at probe area B with $L = 1.88$ mm. The measured optical densities are much higher than those measured in previous pulse photolysis studies with $(3–5)\times10^{10}$ $W/cm^2$ [13, 15], implying that a higher density of solvated electrons is produced by the higher-intensity laser pulse ($4\times10^{13}$ $W/cm^2$) used in this study.

The optical density showed temporal behavior with similar decay timescale and similar response to $NO_3^-$ scavenger in comparison with that of the shadowgraph imaging. To verify the scavenging effect, the decay kinetics of solvated electrons in $KNO_3aq$ were analyzed by calculating the reaction rate of the anionic scavenging reaction with second order kinetics [19, 23], as follows:

$$[e_{aq}^-] = [e_{aq}^-]_0 exp(-k_1[NO_3^-]_0 t). \qquad (2)$$

The data of decay kinetics in $KNO_3aq$ were fitted with exponential decay curves by changing the rate constant $k_1$ as a fitting parameter. The fits are depicted with red and blue solid lines in Fig. 3. In probe area B, the half-lifetime was found to be 170 ps in 0.2 M $KNO_3aq$ and 90 ps in 0.4 M $KNO_3aq$. The rate constant $k_1$ was found to be $2.0\times10^{10}$ $M^{-1}s^{-1}$ in the case of both concentrations. The situation is similar in probe area A, with the lifetime and rate constant being 160 ps and $2.2\times10^{10}$ $M^{-1}s^{-1}$, respectively. The obtained rate constants are in good agreement with the previously reported value of $(0.9–2.2)\times10^{10}$ $M^{-1}s^{-1}$ for scavenging reactions between solvated electrons and $NO_3^-$ [19, 23]. Therefore, the observed temporal dynamics of the shadows and the optical densities should represent the dynamics of solvated electrons.

Next, we discuss the decay kinetics of solvated electrons in pure water. In probe area A, assuming an exponential decay, the 1/e lifetime was estimated to be 900 ps using the initial part of the decay data with a delay $t < 400$ ps, as depicted by the black dashed line in Fig. 3(a). In the early stage of the decay of solvated electrons, geminate recombination is well known to be the loss process of solvated electrons with a short timescale. Geminate recombination is a fast annihilation process by which a solvated electron is pulled back to a parent ion by Coulomb interaction and is recombined with a hydronium counter ion ($H_3O^+$) and another radical such as a hydroxyl (OH) radical, which is often observed on 10–100 ps timescale [12–17]. However, the observed lifetime is much longer than those observed in previous photolysis studies. This unusual decay kinetics could be rationalized by the balance between the loss and production of solvated electrons. The loss of solvated electrons at such a short timescale should be attributed to geminate recombination. On the other hand, in femtosecond laser-induced plasma, free electrons are produced by the initial multiphoton ionization during the duration of a pump pulse at $t \sim 0$ ps. Furthermore, following the passing of the pump pulse through the focal point ($t > 0$ ps), additional free electrons are produced by collisional ionization and photoionization due to the de-excitation of other excited species. Accordingly, additional solvated electrons can be produced even at $t > 0$ ps by the abundant free electrons generated by these ionization processes. Therefore, the production of additional solvated electrons due to the abundant electron density might prolong the duration of geminate recombination process, largely affecting the picosecond to nanosecond dynamics of solvated electrons. In fact, the production of additional solvated electrons can be supported by the further increase in optical density at $t = 0–100$ ps (Fig. 3(a)).

As for probe area B, the 1/e lifetime was estimated to be 500 ps, assuming an exponential decay, as shown by the black dotted line in Fig. 3(b). This lifetime is also much longer than those observed in previous photolysis studies [12–17], but is shorter than that in probe area A. As in the case of probe area A, this unusually longer lifetime could in probe area B be explained by the balance between the loss of solvated electrons by geminate recombination and the additional solvation due to the abundant free electrons generated with laser-induced plasma. In probe area B, the density of free electrons is lower than that in probe area A, as measured below, because the pump pulse intensity in probe area B is lower than in probe area A. The lower free electron density could cause the shorter lifetime of geminate recombination in probe area B because of the smaller degree of additional solvation of free electrons. The laser-intensity-dependence of the decay kinetics of solvated electrons could also support the scenario of the longer lifetime of geminate recombination owing to the additional solvation by abundant free electrons generated with laser-induced plasma.

On the other hand, the exponential decay fits deviated from the experimental decay data with an increase in the probe delay in both probe areas (Fig. 3). This result suggests that the decay kinetics is not explained only by the fast decay in the early stage and the decay reaction rate changes with time. Self-recombination is well known to be another decay process with a longer timescale. In self-recombination, two solvated electrons recombine with each other along with surrounding water molecules [30]. This recombination obeys the reaction rate equation shown below,

$$[e_{aq}^-] = [e_{aq}^-]_0 / (1 + 2k_2 [e_{aq}^-]_0 t) \quad (3)$$

with a rate constant $2k_2$ of $1.1 \times 10^{10}$ M$^{-1}$s$^{-1}$ [31]. The measured densities of solvated electrons in our experiment yield half lifetimes of 300 ns and 6 μs in probe area A and B, respectively, both of which are similar to those observed after $t = 2$ ns. Therefore, self-recombination is considered to be a dominant process in the loss of solvated electrons after $t = 2$ ns, although the decay times are much longer compared to the timescale of the probe delay in our experiment.

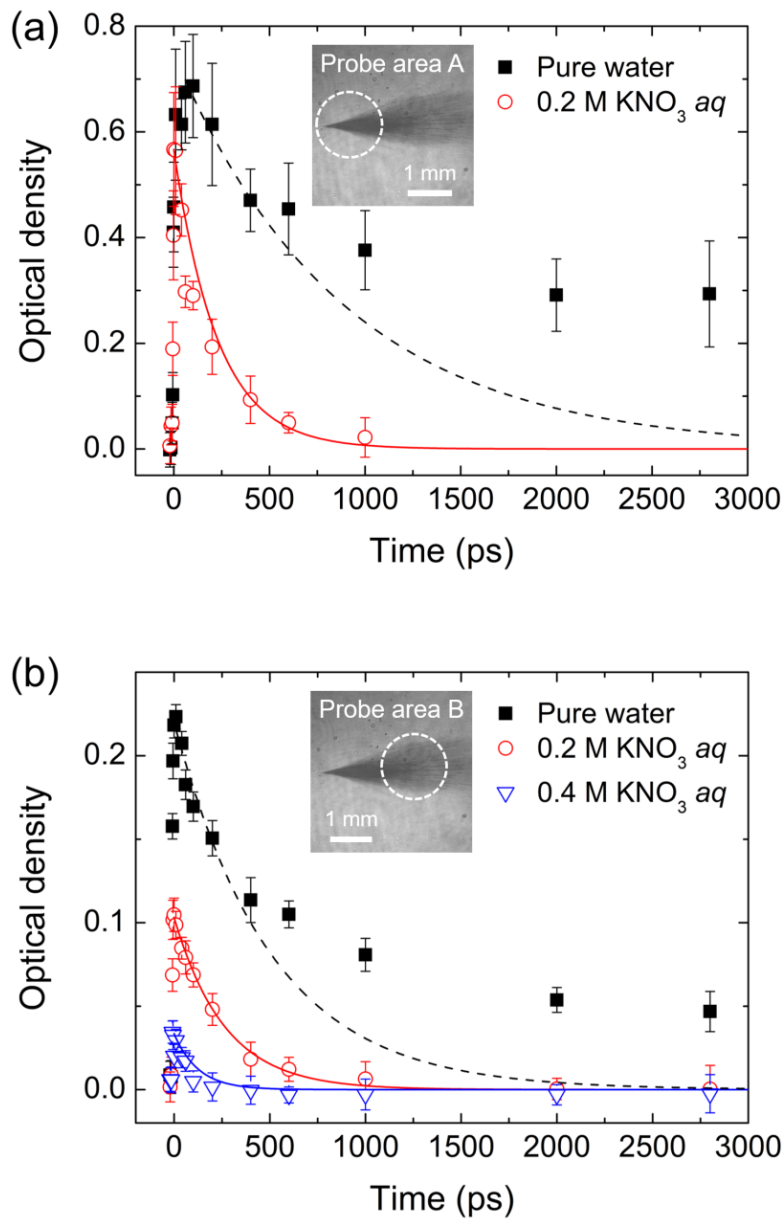

**FIG. 3.** Temporal change in the optical density (**a**) in the vicinity of the focal point (probe area A) and (**b**) in the middle of the focusing cone (probe area B), indicating the decay kinetics of solvated electrons in pure water (black squares) and in 0.2 M (red circles) and 0.4 M (blue triangles) $NO_3^-$ solutions. The inset shadowgraph image shows the position of the probe beam in the plasma at $t = 0$ ps. The red and blue solid lines represent the fits of the decays in the presence of the scavenger. The black dashed line represents the exponential fit for the initial decay kinetics within 400 ps in pure water. Error bars denote the standard deviations of the averaged data, which are caused by the shot to shot variations.

## C. Density of free electrons

To verify the possibility that free electrons influence the decay kinetics of solvated electrons, the density of free electrons in pure water was estimated by interferometry. The interferometry images at $t = 0$ ps are shown in Fig. 4. The electron density ($n_e$) that was line-averaged along the probe beam path was deduced from the phase shift ($\Delta\varphi$) following the generation of plasma [32], which is expressed as,

$$\Delta\varphi(x,y) = \frac{\pi}{\lambda n_c} n_e(x,y) L \qquad (4)$$

where $\lambda$ is the wavelength of the probe pulse, $n_c$ is the critical electron density ($1.75 \times 10^{21}$ cm$^{-3}$ at 800 nm), and $L$ is the path length of the probe pulse as in the case of the absorption measurement. The spatially averaged electron density along the axis of the pump pulse was calculated at $t = 0$ ps to be $1 \times 10^{18}$ cm$^{-3}$ with $L = 0.33$ mm in the probe area A (A in Fig. 4), and $7 \times 10^{16}$ cm$^{-3}$ with $L = 1.35$ mm in the probe area B (B in Fig. 4). The measured free electron densities are much lower than the critical density for the probe pulse wavelength of 800 nm, which confirms that the absorption measurement was not influenced by the absorption and reflection of probe pulse by free electrons. On the other hand, the obtained free electron density is higher than the density of solvated electrons, and the ratio of free electron density to solvated electron density is 1.8–2. The abundance of free electrons could support the hypothesis that free electrons influence the decay kinetics of solvated electrons. Furthermore, the electron density in probe area A is much higher than that in probe area B. The spatially different electron densities could support the laser-intensity-dependence of the prolonged lifetimes of geminate recombination in probe area A and B, in agreement with the scenario of the decay kinetics of solvated electrons.

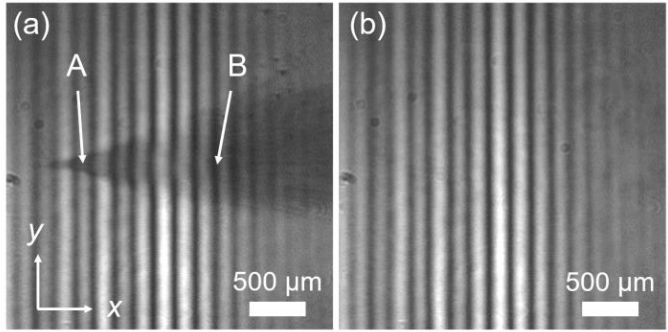

**FIG. 4.** (a) Interferometry image of the plasma in pure water at $t = 0$ ps. A and B in the image indicate the center of probe area A and the left edge of probe area B in Fig. 3, respectively. (b) Interferometry image without the plasma.

**D. Model analysis of the longer decay of solvated electrons**

To further investigate the hypothesis that the unusually longer decay kinetics can be attributed to additional production of solvated electrons, a model calculation of the decay kinetics of solvated electrons was performed. In the model calculation performed here, annihilation of solvated electrons by geminate recombination was expressed by previously reported well-established model by Thomsen et al. [15], and the additional production rate of solvated electrons was added to the model.

In the model of Thomsen et al., total survival probability of solvated electrons for an initial fragment separation of $r_0$, $\Omega(r_0, t)$, is given as a product of each survival probabilities of geminate recombination with $H_3O^+$ and OH radical as shown below, because $H_3O^+$ and OH radical are main recombination partners of solvated electrons in water [14].

$$\Omega(r_0, t) = \Omega_{H_3O^+}(r_0, t)\, \Omega_{OH}(r_0, t). \tag{5}$$

Under the assumption that the recombination of solvated electrons with OH radical is diffusion controlled, survival probability for recombination of solvated electrons with OH radical can be expressed as

$$\Omega_{OH}(r_0, t) = 1 - \frac{a_{OH}}{r_0}\mathrm{erfc}\left\{\frac{r_0 - a_{OH}}{\left[4(D_{e_{aq}^-} + D_{OH})t\right]^{\frac{1}{2}}}\right\}, \tag{6}$$

where $D_{OH} = 2.2 \times 10^{-9}$ m²s⁻¹ is the diffusion constant of OH radical, $D_{e_{aq}^-} = 4.8 \times 10^{-9}$ m²s⁻¹ is the diffusion constant of solvated electron, and $a_{OH} = 0.57$ nm is the reaction radius between OH radical and solvated electrons. On the other hand, survival probability for the recombination of solvated electrons with $H_3O^+$, which is caused by Coulomb attraction, can be approximated by

$$\Omega_{H_3O^+}(r_0, t) = 1 - W_\infty W^*(t) \tag{7}$$

with

$$W_\infty = \frac{1 - \exp(-r_c/r_0)}{1 - \exp(-r_c/a_{H_3O^+})[1 - r_c(D_{e_{aq}^-} + D_{H_3O^+})/(v_{H_3O^+} + a_{H_3O^+}^2)]}, \tag{8}$$

$$W^*(t) = \mathrm{erfc}(B) - \exp(A^2 + 2AB)\mathrm{erfc}(A + B), \tag{9}$$

$$A = \frac{4a_{H_3O^+}^2 + C}{r_c^2}[t/(D_{e_{aq}^-} + D_{H_3O^+})]^{1/2}\sinh^2(-r_c/2a_{H_3O^+}), \tag{10}$$

$$B = r_c \frac{\coth(-r_c/2a_{H_3O^+}) - \coth(-r_c/2r_0)}{4[(D_{e_{aq}^-} + D_{H_3O^+})t]^{1/2}}, \quad (11)$$

$$C = v_{H_3O^+} - \frac{r_c(D_{e_{aq}^-} + D_{H_3O^+})}{a_{H_3O^+}^2 [1 - \exp(r_c/a_{H_3O^+})]}. \quad (12)$$

Here $r_c = e^2/4\pi\varepsilon_0\varepsilon k_B T = 0.712$ nm is the Onsager radius, where $e$ is the elementary charge, $\varepsilon_0$ is the permittivity of vacuum, $\varepsilon = 78$ is the dielectric constant of water, $k_B$ is Boltzmann constant, and $T$ is temperature of water. $D_{H_3O^+} = 9.0 \times 10^{-9}$ m$^2$s$^{-1}$ is the diffusion constant for H$_3$O$^+$, $a_{H_3O^+} = 0.5$ nm is the reaction radius between H$_3$O$^+$ and solvated electrons, and $v_{H_3O^+} = 3.83$ m/s is the reaction velocity. In this calculation, average initial fragment separation distance $<r_0>$ is the only adjustable parameter. The model was validated by the reproduction of previously reported temporal behavior of solvated electrons in Lu et al. [12] with an initial fragment separation distance $<r_0> = 0.9$ nm, as shown in Fig. 5(b) with blue circles and a blue dashed line. The value of $<r_0>$ is in good agreement with the previous studies [12, 15].

The shape of additional production rate of solvated electrons was assumed as shown by G($x$) in Fig. 5(a). Rate of the additional production of solvated electrons were assumed to increase linearly until $t = 100$ ps and subsequently decay exponentially with 1 ns 1/$e$ lifetime. As shown by the black squares in Fig. 3(a), the number density of solvated electrons in pure water decreased once at $t = 40$ ps, and increased again at $t = 100$ ps. This behavior could be attributed to the additional production of solvated electrons from abundant free electrons via such as collisional ionization. Therefore, it is quite reasonable to assume that the temporal variation rate of the additional production of solvated electrons has a short rise time of 100 ps. The result of the model calculation below was not different when the decay lifetime of the additional production rate (1 ns) was changed between 500ps – 5 ns.

The additionally produced solvated electrons were assumed to undergo the same recombination process of the previous model with an average initial fragment separation distance $<r_0> = 0.9$ nm, as soon as they were produced. The total survival probability of solvated electrons considering the additional production, $\Omega_G(r_0, t)$, can be expressed as

$$\Omega_G(r_0, t) = \Omega(r_0, t) + \int_0^t \alpha G(x) \Omega(r_0, t-x) dx, \quad (13)$$

where $\alpha G(x)$ is the additional production rate of solvated electrons at $t = x$, and $\alpha$ is the fitting constant.

The result of the calculation by the modified model is shown in Fig. 5(b), in which $\Omega_G(r_0, t)$ is depicted with a red solid line. The longer decay kinetics of solvated electrons obtained in this study (black squares) was reproduced. Furthermore, the local drop and hill of solvated electrons at $t = 0$–100 ps were also reproduced. Therefore, the modified model calculation demonstrates that there is a proper additional production rate of solvated electrons that can reproduce the longer decay kinetics of solvated electrons with a longer lifetime observed in this study. This result supports the suggested hypothesis that the longer decay kinetics can be attributed to the balance between the loss by geminate recombination and the additional production of solvated electrons by free electrons generated with the laser-induced plasma.

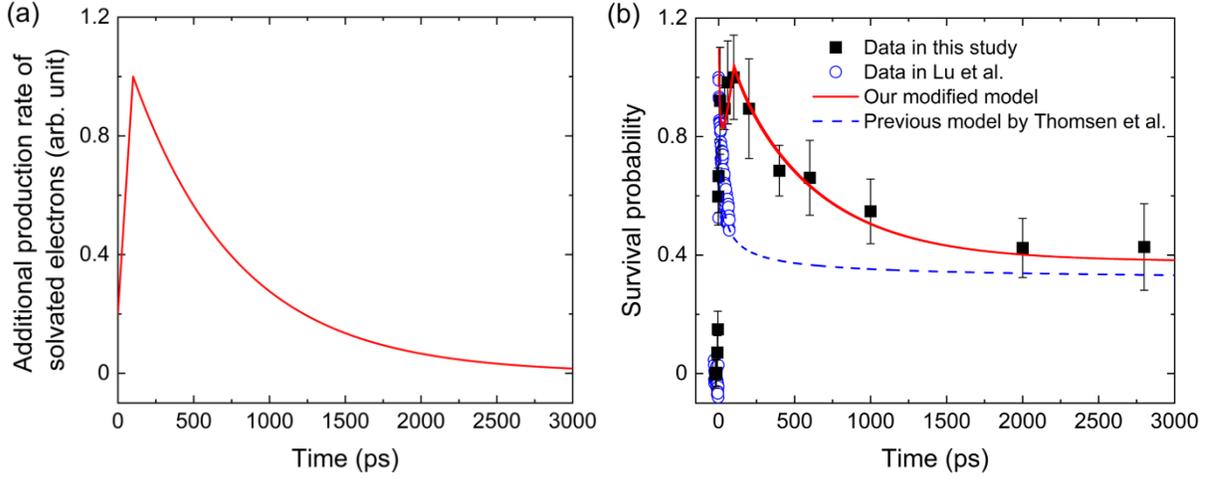

**FIG. 5** Model calculation of decay kinetics of solvated electrons with considering additional production of solvated electrons. (a) Normalized rate of additional production of solvated electrons, G($x$), that was assumed in the calculation. (b) Temporal behavior of solvated electrons with considering the additional production of solvated electrons. Black squares are the normalized experimental data obtained in this study in pure water at probe area A (the data displayed in Fig. 3(a)), blue circles are the previously reported decay data of solvated electrons in Lu et al. [12], red solid line represents the calculated decay of solvated electrons with considering additional production of solvated electrons, $\Omega_G(r_0, t)$, and blue dashed line represents the calculated decay of solvated electron without considering additional production of solvated electrons, $\Omega(r_0, t)$.

## IV. CONCLUSION

We evaluated the dynamics of solvated electrons in the early stage of plasma generation in water with an intense femtosecond laser pulse by a pump-probe technique. Solvated electrons were widely distributed in the focusing cone of the pump pulse, while a plasma plume was initiated in the vicinity of the focal point. The decay kinetics of solvated electrons indicated that geminate recombination occurred with a more prolonged lifetime than that observed in previous pulse photolysis studies. This unusual decay kinetics with a longer lifetime can be attributed to the balance between the loss by geminate recombination and the additional solvation by free electrons generated with the laser-induced plasma. This explanation to the longer decay kinetics was also supported by the modified model calculation of geminate recombination process, by assuming additional production of solvated electrons. Our results imply significant influence of free electrons on the dynamics of solvated electrons during femtosecond laser-induced plasma generation in water. Our study would accelerate further application of femtosecond laser-induced plasma in aqueous media, exploiting the complex reactivity of the plasma abundant in both free and solvated electrons.

**AUTHOUR INFORMATION**

**Corresponding author**

*Email: n.sakakibara@plasma.k.u-tokyo.ac.jp, e-miura@aist.go.jp

**Notes**

The authors declare no competing financial interest.

**ACKNOWLEDGEMENTS**

One of the authors (N.S.) was supported by a Grant-in-Aid via a Japan Society for the Promotion of Science (JSPS) Research Fellowship (Grant No. 19J13045).